# Effects of Work-From-Home on University Students and Faculty

*By Avni Singh*

# Contents

# Acknowledgements

I wrote this paper because of my interest in organisational management and how organisations behave in unprecedented situations. Having learnt through the work-from-home period myself, I know the perspective of many students about learning due to the pandemic but learning the perspectives of university students and faculty was an enlightening experience that could only have happened with the help of some individuals.

I would firstly like to thank Professor HS Jamadagni for guiding me in writing this report and for mentoring me during my internship at the Indian Institute of Science.

Secondly, I would like to thank everyone that I interviewed, including but not limited to: Professor Khincha, Professor Murthy, Professor Prabhakar, Professor Ananthanarayan, and the students at IISc and other universities. Their inputs were paramount to the findings of this paper.

Lastly, I would like to thank my parents, for providing me with the framework required to develop a love for organisational management, and for also making my trip to Bangalore for the internship possible.

# Abstract

The work-from-home policy affected people of all demographics and professions, including students and faculty at universities. After the onset of the COVID-19 pandemic in 2020, institutions moved their operations online, affecting the motivation levels, communication abilities, and mental health of students and faculty around the world. This paper is based mainly on primary data collected from students from around the world, and professors at universities in Bengaluru, India. It explores the effects of work-from-home as a policy in terms of how it changed learning during the pandemic and how it has permanently altered it in a post-pandemic future. Further, it suggests and evaluates policies on how certain negative effects of the work-from-home policy can be mitigated.

# 1. Introduction

The first COVID-19 lockdown in India began in March 2020[1], impacting many strata of society—notably university level students and teachers who had to move their education online. The work-from-home policy implemented worldwide caused teachers to scramble to learn how to navigate the world of online learning, corporations to cease travelling and working in-person, and students to study from home on their laptops.

The work-from-home policy is controversial among both the people it affected and researchers, since it has a variety of advantages and disadvantages. Across the workforce, the work-from-home was advantageous in that it created better job attitudes, increased productivity, employee benefits, workplace diversity, and, in some cases, lower attrition rates. The disadvantages include a lack of motivation, increased distraction, less space for team building, and a sense of isolation.

Two demographics affected significantly by the policy are students and teachers, especially in the higher education sector. During an internship at IISc, I had the opportunity of interviewing samples of these groups and gaining an insight into how the work-from-home policy—and the pandemic, in general—influenced their lives. This paper will investigate the results of these interviews and will contain samples from interviews with specialists in the field of education.

---

[1] "One Year since a Complete Lockdown Was Announced, We Look Back on How India Fought Covid - First Lockdown Announced." The Economic Times, India Times, 24 Mar. 2021, https://economictimes.indiatimes.com/news/india/one-year-since-a-complete-lockdown-was-announced-we-look-back-on-how-india-fought-covid/early-rules-about-masks/slideshow/81662797.cms.

## 2. Background

For university students, the effects of the work-from-home policy were numerous. According to a study conducted in the UK investigating the mental health of university students during the pandemic, the stresses of online education and the restrictions coming along with the pandemic have put university students at greater risk of developing mental health issues[2]. Moreover, the pandemic also gave rise to cheating. A survey by the NCAA in the United States revealed that more than 90% of college students have cheated, in one way or another[3]. These forms of cheating include using unauthorized electronic resources, working together on an individual assignment, paraphrasing without citation, passing someone else's work as their own, or cheating in any way in an exam.

For professors, the pandemic has had several effects of a similar magnitude. During the pandemic, teachers found themselves adapting to teaching from a computer. Their instruction had to be far more engaging to keep the attention of their students, which was a significant challenge, since most of them were not accustomed to computer teaching in the first place. A survey by The Chronicle of Higher Education[4] found out the effects of the pandemic on faculty at universities. A majority of professors faced elevated levels of frustration, anxiety, and stress. They also stated that they struggled with increased workloads and a deterioration of their work-life balance, especially female faculty. More than half of the faculty interviewed stated that they were considering retiring or changing careers, with tenured faculty showing a higher likelihood of retirement than other faculty.

This paper will explore the effects of e-learning and the pandemic on professors and students pursuing their post-graduate or PhD degrees. It will explore the advantages and disadvantages of the work-from-home policy implemented in universities primarily through primary perspectives, with the views supported by secondary information. Lastly, it will explore measures that universities have taken to solve the problems related to work-from-home and analyse how effective these were.

# 3. Methodology

In this report, I will investigate the effects of the pandemic on university students and faculty, with an emphasis on the effects of work-from-home. This will be done vis-à-vis primary and secondary data. The former will be the basis of my analysis of the impact, and the latter will be used to suggest solutions and policies that could change the way work-from-home is integrated into schools and mitigate its effects.

A large portion of my primary data was collected from students and faculty at the Indian Institute of Science (IISc), where I was doing an internship. Here, I interviewed students and faculty using a multiple-choice-question format, with my questions structured in the following format:

- **Before work-from-home**

  This part of the survey was concerning aspects of learning and teaching before the pandemic and how respondents felt about them.

- **During work-from-home**

  This part of the survey asked questions about how significantly the various aspects of learning or teaching mentioned in the pre-work-from-home section were impacted, along with technical questions (which talked about topics like the platforms used for collaboration) and the impact of the pandemic on personal and student life.

- **After work-from-home**

  This part of the survey concerned how the work-from-home has influenced learning and teaching in the offline setting.

The student survey will be Annexure 1, and the faculty survey will be Annexure 2. These surveys were filled out by non-IISc students in an online survey and filled out on a questionnaire sheet by students at IISc. All of the faculty surveys were filled out on a questionnaire sheet, and I recorded any additional comments on a sheet of paper.

I also interviewed two academics from outside of IISc in a less structured setting, although the discussions I had with them pertained to the same topic. I took notes on paper during these interviews and will be recording the notes in the results section of this paper.

My secondary data was collected primarily through online sources, like news articles, studies, and opinion pieces. This data will be used for the section on policy evaluation of this paper, where I will analyse the policies that I had observed from my primary research and see how successful they were.

# 4. Results

## A. Interviews

### i. Professor H.P. Khincha

The first conversational interview I conducted was with professor H.P. Khincha, the retired Vice-Chancellor of Bangalore University. Having taught for over forty years at the Indian Institute of Science (IISc), he has also been associated with the University of Calgary, the National University of Singapore, and EPFL. Currently, he is the Chairman of the Karnataka State Innovation Council.

My interview with Professor Khincha revealed his views on how the pandemic changed the learning landscape, and how it affected his teaching personally. Moreover, I gained an insight into his views on the future of education. In this section of my paper, I will be transcribing my notes from the interview.

Professor Khincha stated that he picked up on digital learning once work-from-home was put into action. However, he felt that the chemistry of learning in person was missing: the interactivity of it, the lack of structure, and the ability to share ideas. He also felt that, because of the meeting recording feature, meetings and classes were more formal. People did not always share their thoughts because they felt they might be misused in a recording. Moreover, Professor Khincha stated that the etiquette of online meetings was very different to in-person meetings, and this changed interactions considerably.

Professor Khincha raised the concern of cheating, stating that, worldwide, 90% of students had cheated in some way, and were not being true to themselves. The first batch of pandemic graduates worried about their performance in the workforce. The first conversion to online education, in Professor Khincha's opinion, was not optimal, and the quality of online education is still subject to improvement.

In the future, teachers will need to work to change the system of education—regarding learning processes and evaluations of learning. The curriculum for online education should also change; there was no time to adapt earlier, which is why this did not happen, but now, it may be beneficial to look at the ways the curriculum could be tweaked to suit online learning.

From a long-term viewpoint, Professor Khincha sees a few things happening to the world of education:

1. Evolution and increases in competitiveness may cause the students in the future to be of a different type and calibre. Professor Khincha spoke of genetically intervened children as well, and how they could impact the future of learning.

2. Robotics in education would change the teaching landscape significantly. A robotic teacher, instead of fitting their teaching style to suit the average child, could adapt its teaching capabilities to each child's needs, thereby making education more accessible for all types of children.

3. He also believes that evaluations will be changed based on the tools available to the learner, and the curriculum should also be changed accordingly.

## ii.   Professor K.N.B. Murthy

The second conversational interview I conducted was with Professor Murthy, the Vice-Chancellor at Dayananda Sagar University in Bangalore. Previously, he was the Vice-Chancellor of PES University in Bangalore and the Director/Principal of PES Institute of Technology. Overall, Professor Murthy has experience of over four decades in teaching, training, research, administration, and industry.

My interview with Professor Murthy revealed a lot about his experience with the pandemic, as well as the experiences of faculty and students at Dayananda Sagar University. My questions mainly involved his experiences during the pandemic, and how the students and faculty were affected by work-from-home. In this section, I will detail the notes from the interview.

Dayananda Sagar University, like other universities in Bangalore, had to stop teaching classes in-person on the 14th of March 2020 due to the state of Karnataka's lockdown. The institutions were unprepared for this and took around a week to find suitable online learning resources. After approximately a month, students started adapting to online learning and attended classes more regularly. Professor Murthy stated that, while the first semester of online learning objectively went well, the teaching and learning were not impactful. He also stated that this was the best possible outcome since most students missed the fundamental interactions of in-person learning—the ability to ask questions and contribute to the class rather than just having professors teach. According to Professor Murthy, roughly 60% of the students at Dayananda Sagar University managed their course load aptly, but the other 40% were not successful. He also stated that the top 20%-30% of students at the university did well before the pandemic and continued to do well during and after. The middle-performance and low-performance students were the ones who suffered the most.

Students were also affected personally. Often, they found learning to be boring, and students unaccustomed to computers had trouble navigating online platforms. Professor Murthy also added that many of them were lonely, but only had their families to talk to and confide in.

Another significant issue was cheating during the exams. Due to the classes being held online, the students at the university also wanted their testing to be done online.

The faculty, by and large, was unhappy with the situation. They did not like working from home since they were not accustomed to online teaching. Aside from learning how to use the platforms in the first place, they also had to make classes more engaging—students often lost interest in the classes and would play video games or music instead. For conventional faculty, especially, the shift to online learning was difficult, with fewer people attending class as well, lowering the morale of the professors. Moreover, the faculty was still unsure of the duration of the work-from-home policy, and this uncertainty also affected their experiences teaching adversely.

Professor Murthy believes that hybrid learning is the way that education will move forward, with students at Dayananda Sagar University visiting the campus on certain days and learning online on others. He also believes that the pandemic has caused significant changes in the way material is taught. He stated that, since the 1990s, learning has been increasingly industry-focused, with a lack of learning for the sake of learning. He believes that the pandemic has only exacerbated this issue.

## B. Surveys

### i. Survey with Professors

For this section of the paper, I interviewed faculty around IISc with a multiple-choice survey that I had designed. I also gave them the option to submit additional comments if they wanted to elaborate on their answers. Part of the faculty that I interviewed were Professor B. Ananthanarayan and Professor T.V. Prabhakar.

The surveys that I designed for students and professors were similar, although some details had been tweaked for the individual survey. My results for professors were more uniform in opinion than those for students, but this could, in part, be due to the smaller sample size of professors as compared to the larger one of students, hence providing a smaller range of opinions.

For this reason, I will not be representing the data in this section in a graphic format, but rather describing the opinions provided by the teachers while adding in their additional commentary.

In terms of what they liked in particular about working before the pandemic, 66.7.7% of the faculty said that they liked meeting students and colleagues in person. 33.3% cited the less structured teaching setting, adding that they preferred having human interaction with their students without the structural constraints placed by online learning. Members of faculty provided different ratings for how the pandemic impacted the aspect of their work that they liked the most. 33.3% gave a rating of 2, 33.3% more stated 3, and the last 33.3% stated 5.

When asked what they did not like about working before the pandemic, 66.7% cited a longer commute time. One member of faculty added that this was especially a problem in Bangalore, where traffic has been getting worse. 33.3% stated that they did not like feeling less productive before the pandemic. On a scale of 1-5, when asked how much this aspect of working had changed during work-from-home, 66.7% of the teachers gave a rating of 5, and 33.3% gave a rating of 4.

When asked what their views were on work-from-home when the pandemic first started, the teachers answered unanimously: they did not like it. When asked if these views changed, however, 66.7% stated that they did while the other 33.3% stated that they did not. Professor Prabhakar explicitly stated that his views on work-from-home remained largely positive because of the positive learning outcomes it had for his own work.

When asked to rate how work-from-home impacted their personal lives on a scale of 1-5, the teachers, once again, had different answers. 33.3% answered 1, 33.3% answered 2, and 33.3% answered 5. When asked how it affected their students' lives, the faculty gave differing answers as well. 33.3% rated it 3, 33.3% rated it 4, and 33.3% rated it 2.

In terms of platforms used for collaboration, the faculty at IISc primarily used Teams (60%), Google meet (20%) and Zoom (20%). As for why these platforms were used, the teachers cited convenience, security, and pricing. These platforms stayed the same over the course of the pandemic.

In terms of how the faculty ensured teamwork with work-from-home, all of them answered that they had regular meetings with students and other faculty. Today, 66.7% stated that they were teaching offline only, while 33.3% stated that they had hybrid learning.

When asked to rate their students' motivation while working from home on a scale of 1-5, most teachers rated it on the lower end of the spectrum (below 3, 66.7%), while 33.3% gave a rating of 5. When asked if the pandemic caused more diversity in the workforce/student-force, the teachers answered 'no'. 66.7% of teachers also stated that their classrooms' designs had changed during work-from-home, while 33.3% stated that the design had not changed. Only 66.7% needed the lab for their work, and they said that they still visited the lab.

### iii.  Survey with Students

My survey with students was carried out to identify how students were affected by the work-from-home policy during the pandemic. It spanned 20 questions, in a pre-pandemic, during-pandemic, and post-pandemic structure. The results are as follows.

My survey with students spanned a variety of geographic regions—ranging from India to Australia, to South Korea, to the United States and Canada—since I wanted to interview a more extensive range of students and understand their experiences with work-from-home, which was implemented in all countries affected by the pandemic.

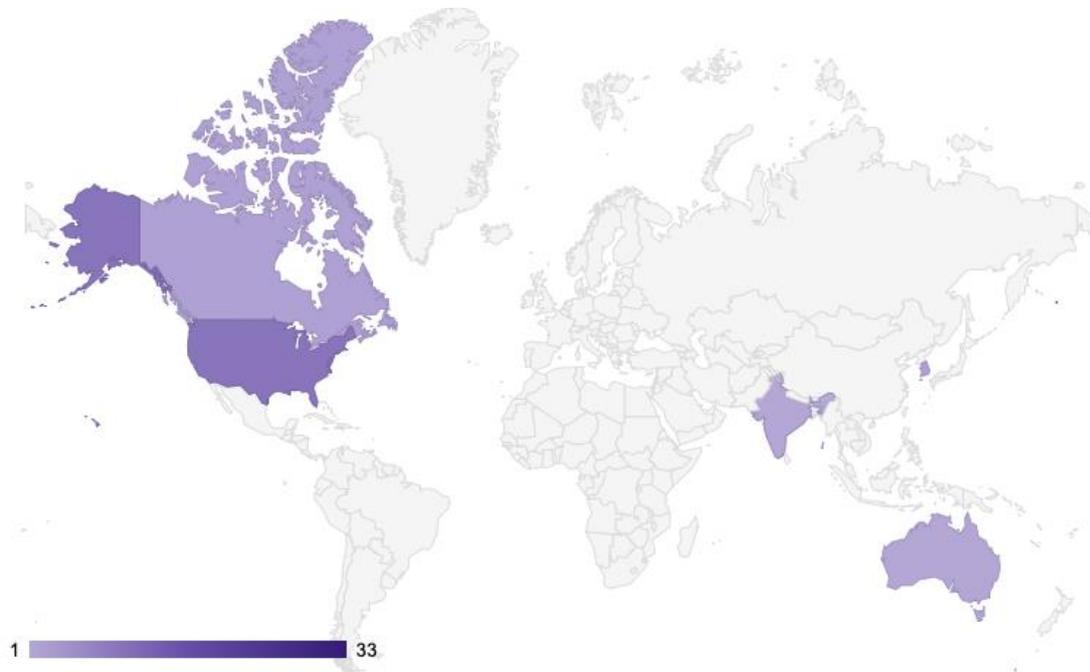

59.2% of students were pursuing their undergraduate degree, along with 24.5% pursuing their

PhDs and 16.3% doing their master's degrees:

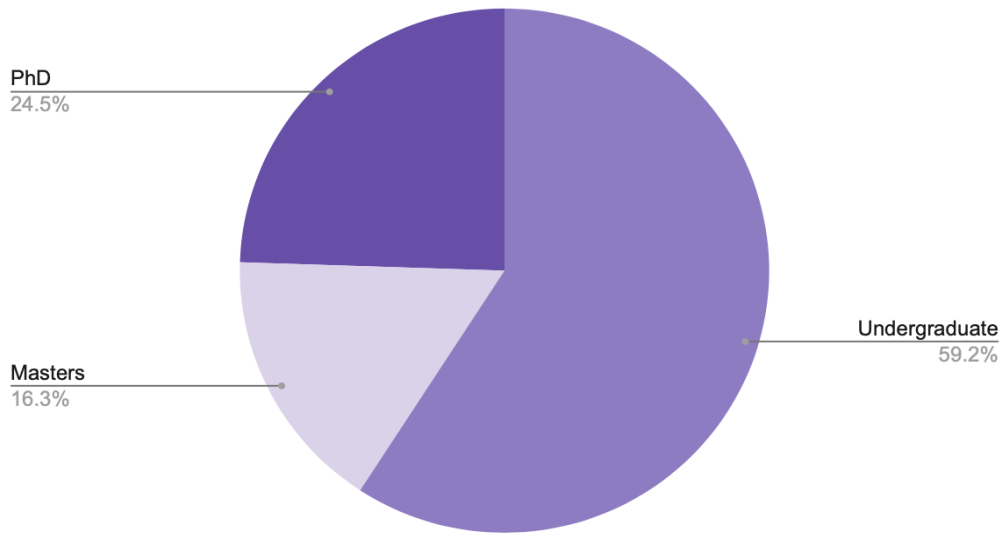

Out of the respondents, the largest percentage of people answered, "meeting students and professors in person" in response to the question, "What did you like in particular about learning before the pandemic?" Consequently, 29.9% answered with "sense of routine", 27.4% answered "increased productivity, 11.1% answered "the less structured setting", and lastly, 0.9% answered "less screen time."

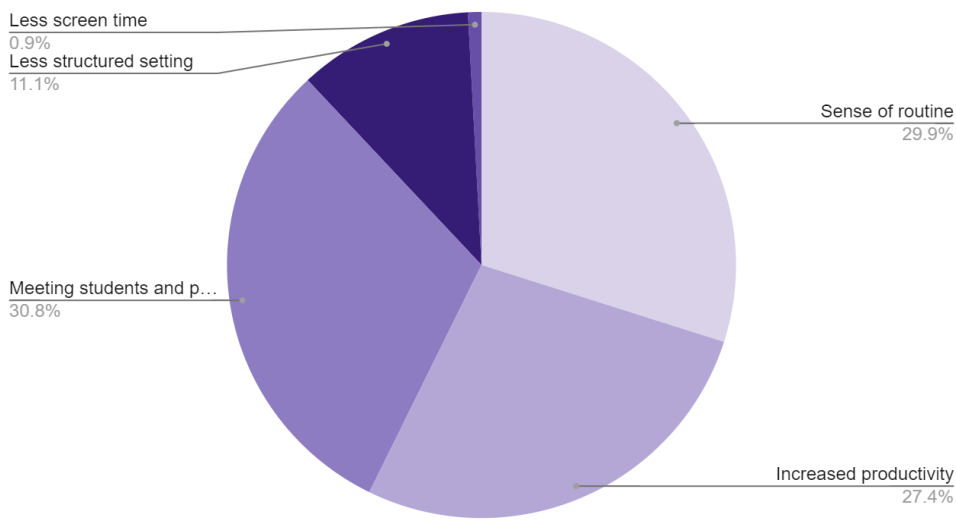

In the next question, when asked if this aspect of learning was impacted by work-from-home—and how significantly, on a scale of 1-5—a vast majority of students answered that it was impacted very significantly (4+ ratings were given by 89.8% of the respondents). In comparison, the other 10.2% answered either 2 or 3. The mean for this question was a score of 4.31, while the mode and median were scores of 4.

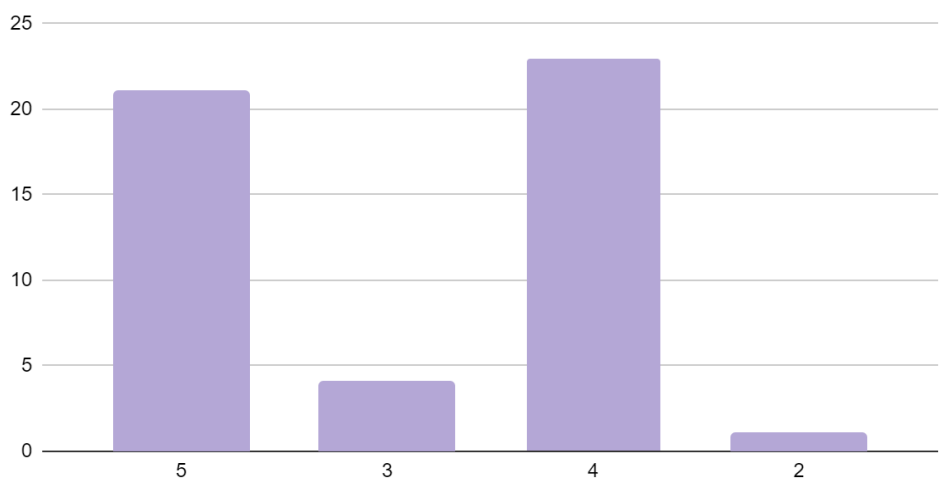

How significantly do you think the pandemic impacted this aspect of your learning, on a scale of 1-5?

The next question asked students what they did not like about learning before the pandemic. 43% answered that it was the fact that they had a less flexible schedule. 27.8% answered that there was a longer commute time. 26.6% said that this was the lack of personal time. 1.3% answered that there were no recordings of lectures to go over and that they were less productive, respectively.

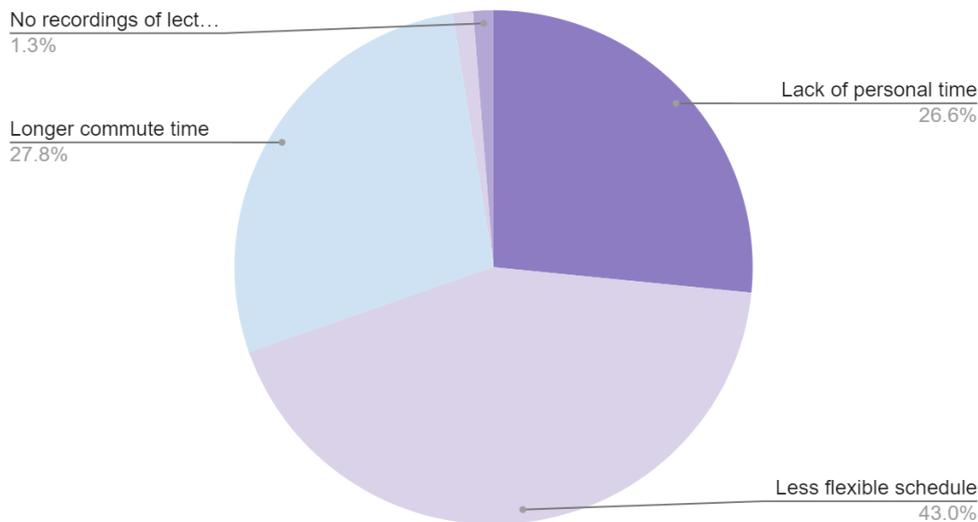

When asked how significantly the pandemic affected this aspect of learning on a scale of 1-5, 46.9% of students answered 4. 22.4% answered 3, and 16.3% answered 5. The cumulative frequency for 1 and 2 is 14.2%. For this question, the mean was a rating of 3.63, the median was a rating of 4, and the mode was also a rating of 4.

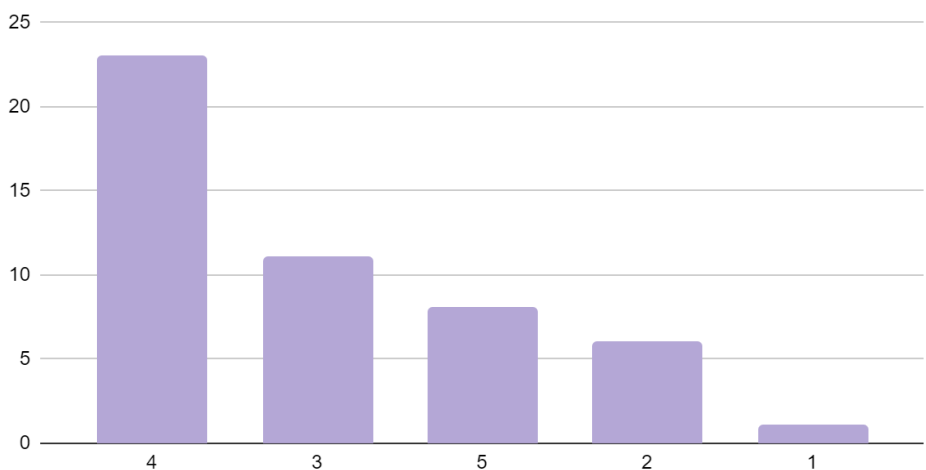

Next, the questions moved on to the pandemic period, where work-from-home was actually in action. The first question here was "What were your views on work-from-home when the pandemic first started?" 42.9% of students felt neutral about work-from-home, while 24.5% disliked it. A 32.7% share of responses was occupied by people who liked work-from-home at the onset of the pandemic.

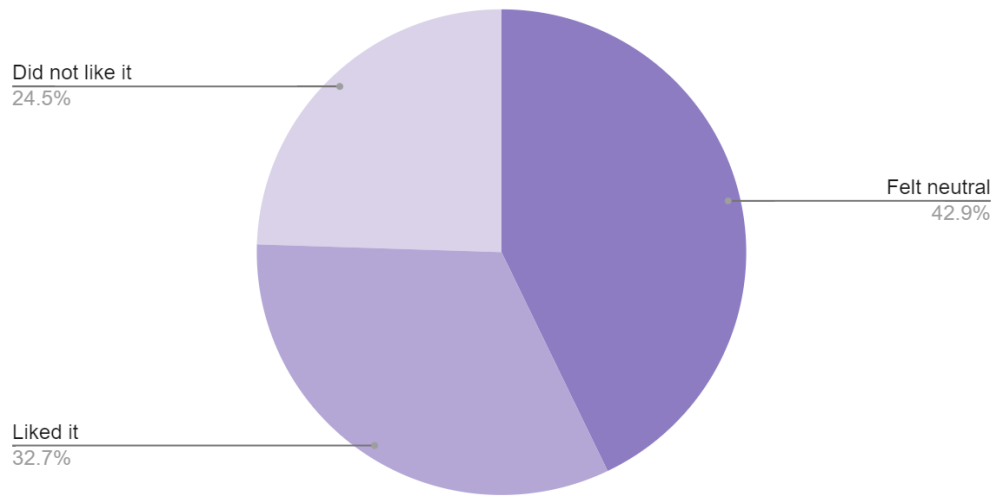

The following question asked whether these views had changed over time. Respondents answered in the following way:

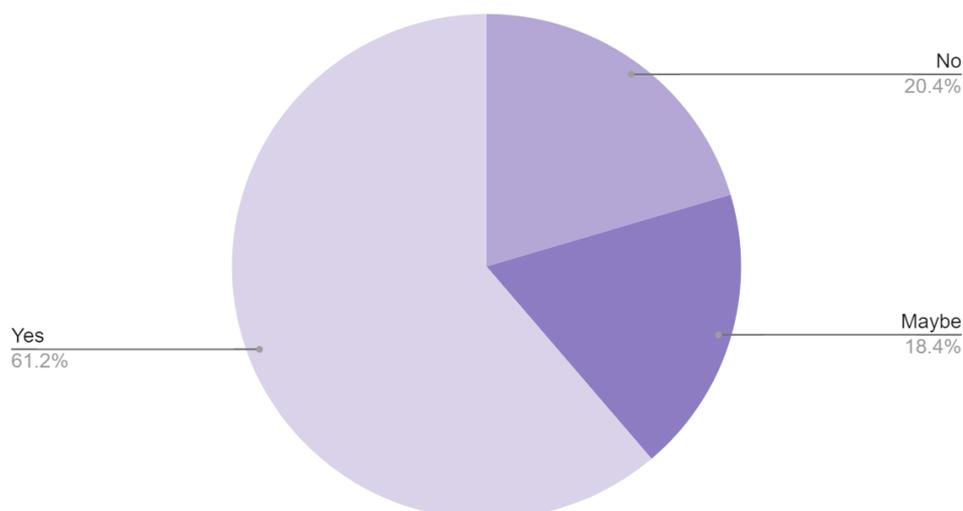

When asked to rate how much work-from-home impacted their personal lives, the respondents gave the following scores, with 30.6% of respondents rating it 4, 32.7% rating it 5, 14.3% rating it 2, and 22.4% rating it 3. Ratings of 1 were not considered by any of the respondents. The mean for this question was 3.81, the mode 5, and the median 4.

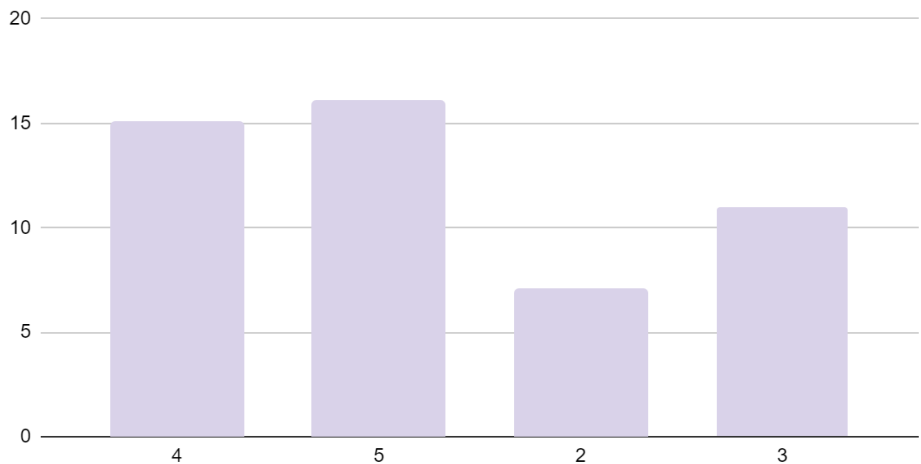

The next question dealt with how work from home impacted the respondents' school lives, on a scale of 1 to 5. 44.9% of the respondents rated this a 5, while 32.6% rated it a 4. 18.4% of respondents rated it a 3, while 4.1% rated it a 2. Ratings of 1 were not considered by any of the respondents. The mean for this question was a score of 4.18, while the mode was 5 and the median was 4.

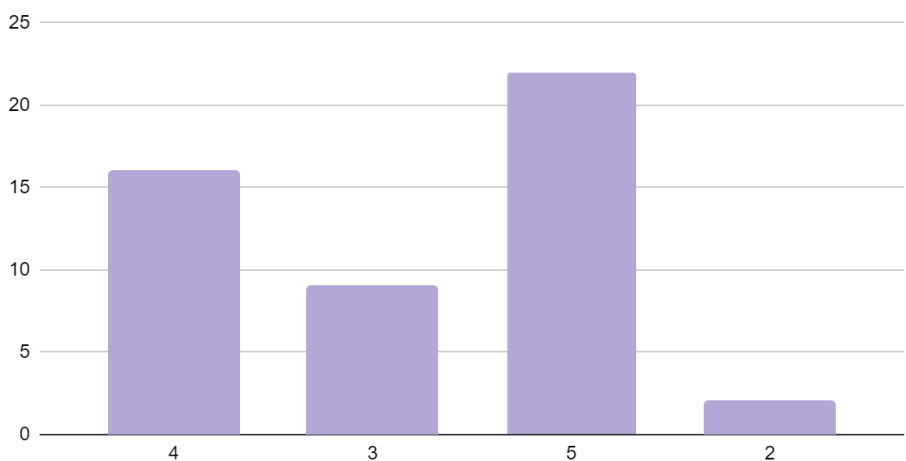

The next question asked students which online-learning platforms they utilised. A vast majority had used Zoom (38.8%), as well as Google Meet (28.6%) and Teams (27.6%). 4.1% of students also used Skype, and 1% used Webex. When asked why their schools used these platforms, 92% of the respondents answered that it was for convenience, with 4.2% citing a superior user interface as the reason, and the last 3.8% saying they were unsure of the reasoning behind the usage of these platforms.

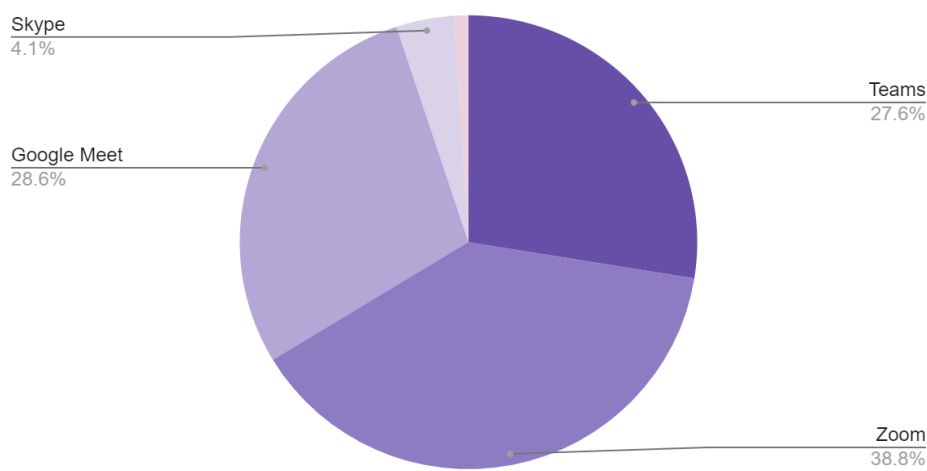

The respondents were also asked how they ensured teamwork during virtual learning. This was an important question to the study because teamwork and human interaction are known to increase motivation (which is what the next question of this survey deals with). The bar chart below shows the results of this question, with a vast majority (64%) of students saying that they kept collaborating through regular meetings, and 34% saying they sent real-time updates to students and professors. 20% held team-building activities, while 6% had collaboration workshops. 2% of respondents made a text group, while 2% also stated that there was no collaboration. With this question, it is important to note that multiple boxes could be ticked, and hence some respondents clicked on more than one answer.

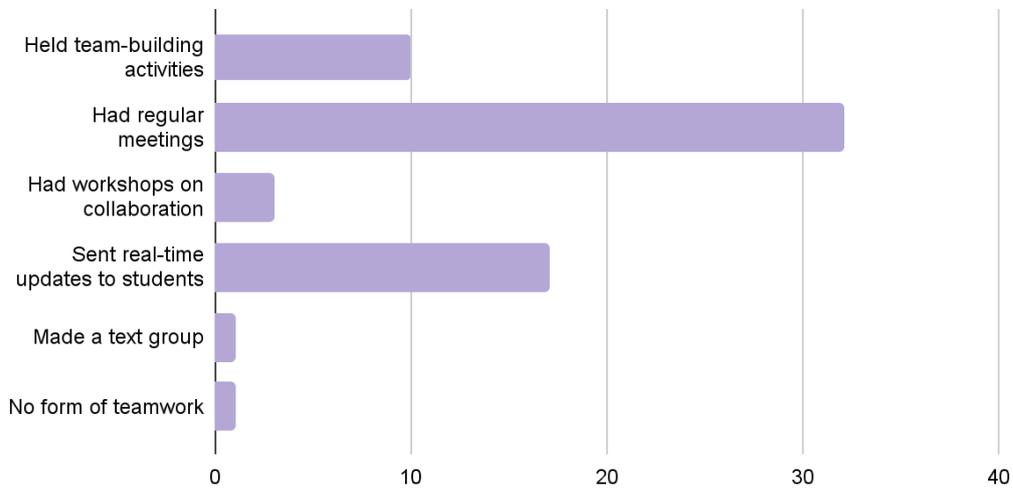

The next question that the survey asked was about motivation during work-from-home. On a scale of 1-5, 1 being the least motivated, most respondents showed a tendency towards being less motivated (with 73.7% rating their motivation 3 and below). The mode and median for this question were ratings of 3, while the mean was a rating of 2.55.

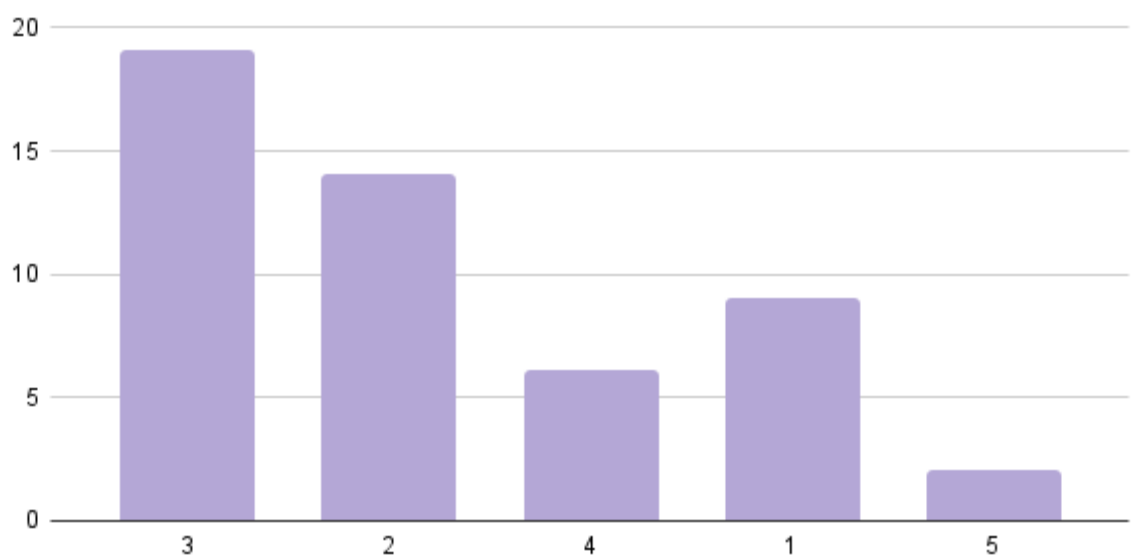

The next question moved the survey into the future: what happened after the work-from-home policy had started to ease off? The respondents were asked whether they were having offline, hybrid, or virtual classes, with the following answers:

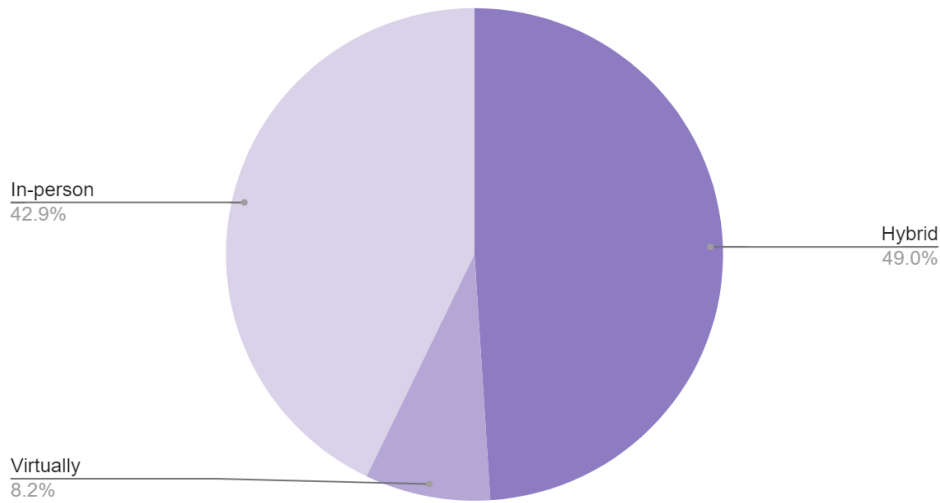

When asked about the work-from-home policy changing the levels of diversity at their school, a majority of the respondents said that it had not caused more diversity.

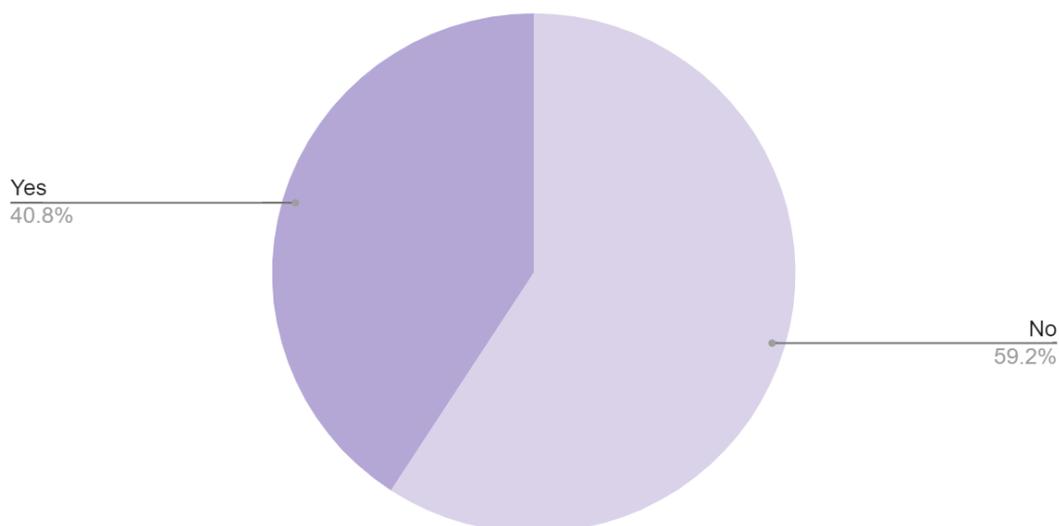

Further, respondents were asked about physical changes to their universities--i.e. whether the work-from-home policy and pandemic had changed anything about their classrooms, such as classroom design. The answers to this question saw the same split-up as the previous question, as shown below:

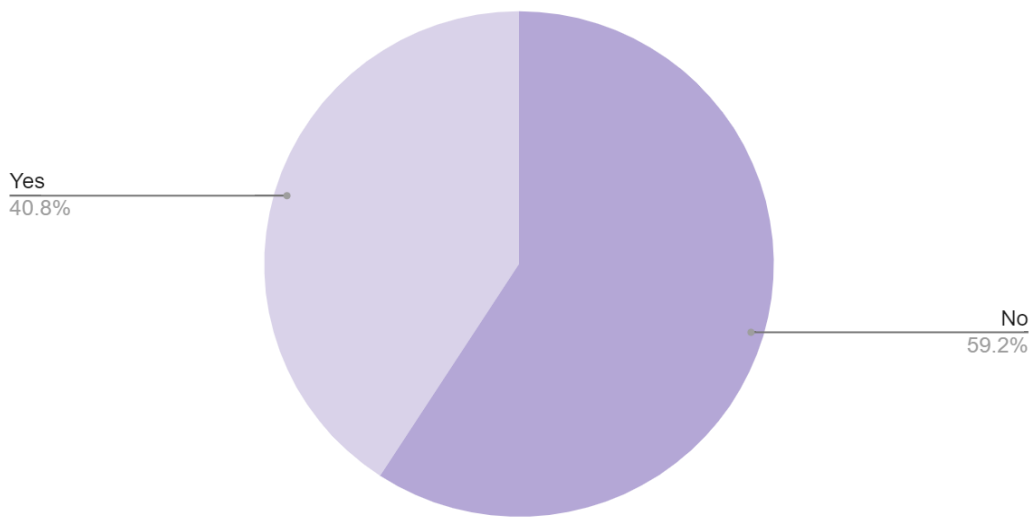

The last question was solely for students who required the lab for their studies. They were asked if they still used the labs at their universities. The answers are shown below:

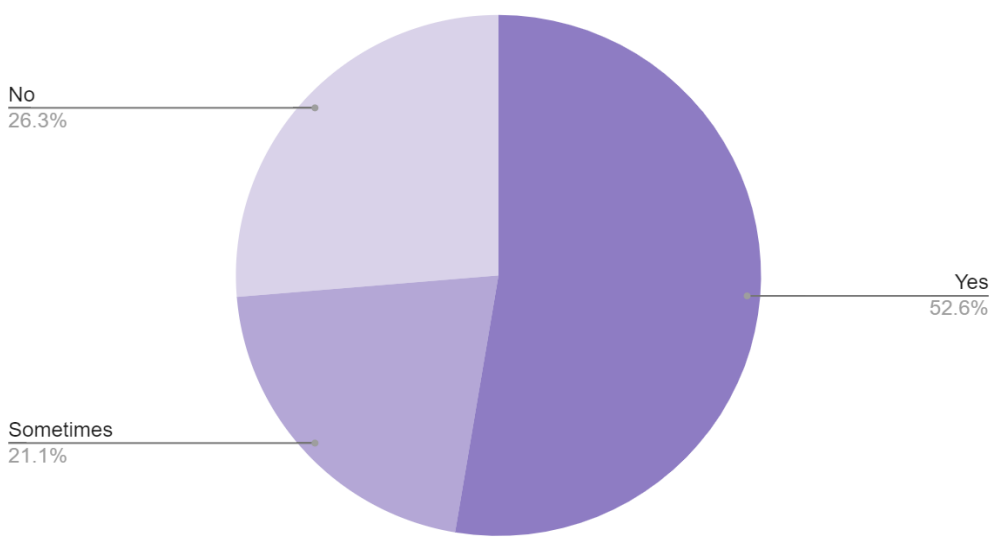

# 5. Analysis of Results

Overall, based on my results for teachers, a reasonable conclusion can be made that, while they may not have enjoyed teaching from home, they were not affected severely, especially in terms of job satisfaction and security. Moreover, they seemed to be more concerned about the welfare of their students--especially in reference to my interview with Professor Murthy. Most of the professors, especially Professor Khincha, saw the world of education changing a lot in the future, moving towards a more hybrid model with the type of students being taught changing drastically as well.

My results for students showed more varied results. While most of them liked learning before the pandemic—meeting teachers and other students, the sense of routine, increased productivity, and the less structured setting being the most significant aspects of learning for them—they also had aspects of learning prior to the pandemic that they disliked—like the longer commute time, less flexible schedule, and lack of personal time. Invariably, all of these changed significantly. The aspects of learning that the students liked changed by an average score of 4.31 out of 5. The aspects of learning that they did not like changed less drastically, but significantly nonetheless, with an average score of 3.63.

They also mostly felt neutral about work-from-home during the pandemic, although a significant chunk of students had strong opinions. Moreover, most of them changed their views over time, although for select portions, they remained the same or did not change much.

Most students also believed that work-from-home impacted their personal lives significantly, judging by the average score of 3.81. The same goes with their student lives, which changed even more, by an average of 4.18. This correlates with the faculty interviews—like, for example, what Professor Murthy mentioned about students feeling less motivated to study and isolating themselves from friends and family during work-from-home.

The responses suggested that the most common form of collaboration during the pandemic was regular meetings—for both students and professors.

An important finding was motivation levels. The levels of motivation were, on average, 2.55 out of 5. This is a low amount, especially considering the median and mode were also very low, at 3 out of 5. This may also correlate with the findings in the previous questions: maybe solely having regular meetings is not enough to ensure collaboration.

Currently, learning is happening in a hybrid or in-person manner for 91.8% of students, and 73.7% of students who have lab-work in their syllabi have started to visit the lab to some extent for their work. This shows that the world is gradually returning to the "new normal", with an emphasis on "new", because 40.8% of students stated that work-from-home has caused more diversity at their schools. On top of this, 40.8% of students also stated that the pandemic caused their classrooms to change in design. This was evident during my internship as well, with the labs having segregated workstations to comply with social distancing rules.

# 6. Mitigating the Negative Effects of Work-From-Home

The COVID-19 pandemic may have died down, but there is no predicting the possible conditions that may arise in the future and affect educational institutions, forcing them to return to a work-from-home format. My survey identified some shortcomings of work-from-home in its current form, so in this section, I will be suggesting and evaluating possible solutions to the issues that arise from the policy.

There is no one-size-fits-all approach to mitigating the negative effects from work-from-home, so this section of my paper will analyse multiple solutions to the different problems identified.

## A. Increasing Interaction

While students and faculty are ready to rid themselves of the longer commute times associated with in-person learning, they are not as prepared to let go of the positive aspects of it: like in-person interaction and less flexible schedules. A possible solution to the problem of a lack of in-person interaction during work-from-home could be to have a set timing for bonding with classmates and professors. While this still relies on the structure that a lot of people would prefer to be without, it enables students and professors to get to know each other better and therefore form connections similar to those formed in an in-person setting. Interaction is especially important in learning, studies find, since it can be effective in organising thoughts, reflecting on understandings from the material learnt, and finding gaps in reasoning.[5]

Another possible solution would be to enable the creation of less formal text groups where classmates could get to know each other. While online friendships are not the same as those formed by real interaction, they can still be effective in mitigating loneliness in students[6].

---

## B. Increasing Motivation

Another significant cause for worry from my survey was the dip in motivation levels for students. While professors remained motivated during the pandemic, according to their scores, students remained less motivated, their levels averaging 2.55 out of 5. Even in the interviews conducted, faculty stated that most students were not motivated to attend classes, and that most of them even cheated.

To tackle motivation, collaboration—mentioned below—is essential. Moreover, they could be encouraged to take part in extra-curricular activities rather than just their academics to break the monotony of a life at home. A more interactive learning session could also encourage motivation in students. For example, allowing students to volunteer information about the topic at hand might make the lesson feel more personalised. During my interview with Professor Prabhakar, he mentioned that teaching interactively was a big challenge but ultimately a fruitful one, since he believes it is incredibly important to hold his students' attention.

## C. Increasing Collaboration

Collaboration was also mentioned in my survey, and while most students and professors cited having regular meetings as their means for collaboration, this may not be enough to produce meaningful collaborative work and interaction. Collaboration is incredibly important: not just in the educational world, but also in business[7].

Similar to part A of this section, collaboration could be encouraged by spending time interacting with students and professors—a way to do this would be to add in specific timings for interaction. Furthermore, group projects and group presentations could be an interesting way to encourage collaboration.

---

[7] Gardner, Heidi K, and Ivan Matviak. "7 Strategies for Promoting Collaboration in a Crisis." Harvard Business Review, Harvard Business Publishing, 1 Feb. 2021, https://hbr.org/2020/07/7-strategies-for-promoting-collaboration-in-a-crisis.

## D.  Bettering Mental Health

The pandemic saw the deterioration of students' mental health, since a lot of them faced difficult home situations and increasing loneliness due to a lack of interaction. Aside from solving these problems indirectly (i.e. encouraging interaction or extra-curricular activities among students), universities could also implement counselling services. This would give the students a judgement-free space to express their feelings and can be executed easily online as well.

# 7. Conclusion

Based on my interviews and surveys, a situation warranting work-from-home is predicted to occur again, and most faculty predict that this will be a better opportunity since schools will have the opportunity to prepare themselves for online learning. In this case, they would be more likely to implement mental health resources and have better collaboration opportunities.

A variety of improvements can be made to schools' online learning programs to encourage motivation, collaboration, and interaction, and improve mental health. This includes creating special timings for students and professors to bond, encouraging the creation of text groups, and encouraging students to become familiar online.

As situations change, the format of work-from-home settings will also change. This paper has analysed and suggested some of the ways it may change in the future, based on the results obtained from interviewing faculty at universities and surveying students around the world.

In conclusion, the work-from-home policy implemented during the COVID-19 pandemic has had a variety of effects on both students and faculty at IISc and other institutions, both negative and positive, and there are many ways to mitigate the negative situations that may arise if work-from-home is implemented again.

# 8. Annexures

## i. Survey for Students

1. Where are you from?

2. What degree are you pursuing?

    a. Undergraduate

    b. Masters

    c. PhD

    d. Other…

3. What did you like in particular about learning before the pandemic?

    b. Meeting students and professors in person

    c. The less structured learning setting

    d. Increased productivity

    e. The sense of routine

4. How significantly do you think the pandemic impacted this aspect of your learning, on a scale of 1-5?

    a. 1

    b. 2

    c. 3

    d. 4

    e. 5

5. What did you not like about learning before the pandemic?

    a. Longer commute time

    b. Lack of personal time

    c. A less flexible schedule

    d. Less productivity

6.  How significantly do you think the pandemic impacted this aspect of your learning, on a scale of 1-5?

    a.  1

    b.  2

    c.  3

    d.  4

    e.  5

7.  What were your views on learning-from-home at the beginning of the pandemic?

    a.  Did not like it

    b.  Liked it

    c.  Felt neutral

8.  Did these views change over time?

    a.  Yes

    b.  No

    c.  Maybe

9.  How do you think learning-from-home impacted your personal life, on a scale of 1-5 (1 being the least significantly, 5 being the most)?

    a.  1

    b.  2

    c.  3

    d.  4

    e.  5

10. How do you think learning-from-home impacted your student life, on a scale of 1-5 (1 being the least significantly, 5 being the most)?

    a.  1

    b. 2

    c. 3

    d. 4

    e. 5

11. What platforms did your school use for collaboration during learning-from-home?

    a. Teams

    b. Zoom

    c. Webex

    d. Google Meet

12. Did your school have a particular reason for using these platforms?

    a. The convenience

    b. Pricing

    c. Easy user interface

    d. Security

13. Did your school change the platforms over the course of the pandemic?

    a. Yes

    b. No

14. How did you ensure teamwork with everything happening virtually?

    a. Held team building activities

    b. Had regular meetings

    c. Had workshops on collaboration

    d. Sent real time updates to students and professors

15. Is learning currently happening completely virtually, in a hybrid manner, or completely offline?

    a. Virtually

      b. Hybrid

      c. Offline

16. How motivated were you and your fellow students, working from home, on a scale of 1-5?

      a. 1

      b. 2

      c. 3

      d. 4

      e. 5

17. Has work-from-home caused more diversity? This could be in terms of gender, geographical location, etcetera.

      a. Yes

      b. No

      c. Maybe (please elaborate)

18. Have the pandemic and the learning-from-home policy changed anything about your physical workplace (like classroom design)?

      a. Yes

      b. No

      c. Maybe (please elaborate)

19. If learning involves lab work for you, do you still visit? Or do you conduct simulations remotely?

      a. I still visit

      b. I conduct simulations virtually

## ii.     Survey for Professors

1. What did you like in particular about working before the pandemic?

    a.  Meeting students and colleagues in person

    b.  The less structured teaching setting

    c.  Increased productivity

    d.  The sense of routine

2. How significantly do you think the pandemic impacted this aspect of your work, on a scale of 1-5?

    a.  1

    b.  2

    c.  3

    d.  4

    e.  5

3. What did you not like about teaching before the pandemic?

    a.  Longer commute time

    b.  Lack of personal time

    c.  A less flexible schedule

    d.  Less productivity

4. Do you think the pandemic substantially impacted this aspect of your teaching?

    a.  Yes

    b.  No

    c.  Maybe (elaborate)

5. What were your views on teaching-from-home at the beginning of the pandemic?

    a.  Did not like it

    b.  Liked it

      c. Felt neutral

6. Did these views change over time?

      a. Yes

      b. No

      c. Maybe

7. How do you think teaching-from-home impacted your personal life, on a scale of 1-5 (1 being the least significantly, 5 being the most)?

      a. 1

      b. 2

      c. 3

      d. 4

      e. 5

8. How do you think teaching-from-home impacted your student life, on a scale of 1-5 (1 being the least significantly, 5 being the most)?

      a. 1

      b. 2

      c. 3

      d. 4

      e. 5

9. What platforms did your school use for collaboration during teaching-from-home?

      a. Teams

      b. Zoom

      c. Webex

      d. Google Meet

10. Did your school have a particular reason for using these platforms?

  a. The convenience

  b. Pricing

  c. Easy user interface

  d. Security

11. Did your school change the platforms over the course of the pandemic, or did they stay the same?

  a. The platforms were changed

  b. They stayed the same

12. How did you ensure teamwork with everything happening virtually?

  a. Held team building activities

  b. Had regular meetings

  c. Had workshops on collaboration

  d. Sent real time updates to students and professors

13. Is teaching currently happening completely virtually, in a hybrid manner, or completely offline?

  a. Virtually

  b. Hybrid

  c. Offline

14. How motivated were you and your fellow students, working from home, on a scale of 1-5?

  a. 1

  b. 2

  c. 3

  d. 4

  e. 5

15. Has work-from-home caused more diversity? This could be in terms of gender, geographical location, etcetera.

    a.  Yes

    b.  No

    c.  Maybe (please elaborate)

16. Have the pandemic and the teaching-from-home policy changed anything about your physical workplace (like classroom design)?

    a.  Yes

    b.  No

    c.  Maybe (please elaborate)

17. If teaching involves lab work for you, do you still visit? Or do you conduct simulations remotely?

    a.  I still visit

    b.  I conduct simulations virtually